\documentclass[11pt,twoside]{article}
\usepackage{./asp2014}
\usepackage{graphicx}
\aspSuppressVolSlug
\resetcounters

\begin{document}

\title{The Gaia Mission, Binary Stars and Exoplanets}
\author{Laurent~Eyer,$^1$ Lorenzo~Rimoldini,$^1$ Berry~Holl,$^1$ Pierre~North,$^2$ Shay~Zucker,$^3$ Dafydd~W.~Evans,$^4$ Dimitri~Pourbaix,$^5$ Simon~T.~Hodgkin,$^4$  William~Thuillot,$^6$ Nami~Mowlavi$^1$ and Benoit~Carry$^6$
\affil{$^1$D\'{e}partement d'Astronomie, Universit\'{e} de Gen\`{e}ve, 1290 Versoix, Switzerland; \email{Laurent.eyer@unige.ch}}
\affil{$^2$Laboratoire d'astrophysique, Ecole Polytechnique F\'{e}d\'{e}rale de Lausanne (EPFL), Observatoire de Sauverny, 1290 Versoix, Switzerland;}
\affil{$^3$Department of Geosciences, Raymond and Beverly Sackler Faculty of Exact Sciences, Tel Aviv University, 6997801 Tel Aviv, Israel;}
\affil{$^4$Institute of Astronomy, University of Cambridge, Cambridge, CB3 0HA, United Kingdom;}
\affil{$^5$FNRS, Institut d'Astronomie et d'Astrophysique, Universit\'{e} Libre de Bruxelles, 1050 Bruxelles, Belgium;}
\affil{$^6$Institut de m\'{e}canique c\'{e}leste et de calcul des \'{e}ph\'{e}m\'{e}rides, Observatoire de Paris,  75014 Paris, France}}

\paperauthor{Laurent~Eyer}{Laurent.Eyer@unige.ch}{ORCID_Or_Blank}{Universit\'{e} de Gen\`{e}ve}{D\'{e}partement dÕAstronomie}{Versoix}{State/Province}{1290}{Switzerland}
\paperauthor{Lorenzo~Rimoldini}{Lorenzo.Rimoldini@unige.ch}{ORCID_Or_Blank}{Universit\'{e} de Gen\`{e}ve}{D\'{e}partement d'Astronomie}{Versoix}{State/Province}{1290}{Switzerland}
\paperauthor{Berry~Holl}{Berry.Holl@unige.ch}{ORCID_Or_Blank}{Universit\'{e} de Gen\`{e}ve}{D\'{e}partement dÕAstronomie}{Versoix}{State/Province}{1290}{Switzerland}
\paperauthor{Pierre~North}{Pierre.North@epfl.ch}{ORCID_Or_Blank}{Ecole Polytechnique F\'{e}d\'{e}rale de Lausanne}{Laboratoire d'astrophysique}{Versoix}{State/Province}{1290}{Switzerland}
\paperauthor{Shay~Zucker}{shayz@post.tau.ac.il}{ORCID_Or_Blank}{Tel Aviv University}{Department of Geosciences}{Tel Aviv}{State/Province}{6997801}{Israel}
\paperauthor{Dafydd~W.~Evans}{dwe@ast.cam.ac.uk}{ORCID_Or_Blank}{University of Cambridge}{Institute of Astronomy}{Cambridge}{State/Province}{CB3 0HA}{United Kingdom}
\paperauthor{Dimitri~Pourbaix}{pourbaix@astro.ulb.ac.be}{ORCID_Or_Blank}{Universit\'{e} Libre de Bruxelles}{Institut d'Astronomie et d'Astrophysique}{Bruxelles}{State/Province}{1050}{Belgique}
\paperauthor{Simon~T.~Hodgkin}{sth@ast.cam.ac.uk}{ORCID_Or_Blank}{University of Cambridge}{Institute of Astronomy}{Cambridge}{State/Province}{CB3 0HA}{United Kingdom}
\paperauthor{William~Thuillot}{William.Thuillot@obspm.fr}{ORCID_Or_Blank}{Observatoire de Paris}{Institut de m\'{e}canique c\'{e}leste et de calcul des \'{e}ph\'{e}m\'{e}rides}{Paris}{State/Province}{75014}{France}
\paperauthor{Nami~Mowlavi}{Nami.Mowlavi@unige.ch}{ORCID_Or_Blank}{Universit\'{e} de Gen\`{e}ve}{D\'{e}partement dÕAstronomie}{Versoix}{State/Province}{1290}{Switzerland}
\paperauthor{Benoit~Carry}{bcarry@imcce.fr}{ORCID_Or_Blank}{Observatoire de Paris}{Institut de m\'{e}canique c\'{e}leste et de calcul des \'{e}ph\'{e}m\'{e}rides}{Paris}{State/Province}{75014}{France}

\begin{abstract}
On the 19th of December 2013, the Gaia spacecraft was successfully launched by a Soyuz
rocket from French Guiana and started its amazing journey to map and characterise
one billion celestial objects with its one billion pixel camera. In this presentation, we
briefly review the general aims of the mission and describe what has happened since
launch, including the Ecliptic Pole scanning mode. We also focus especially on binary
stars, starting with some basic observational aspects,
and then turning to the remarkable harvest that Gaia is expected to yield for these objects.
\end{abstract}

\section{Introduction}
On December 19, 2013 at 10h18 (CET), in French Guiana and in different institutes across Europe, many scientists, software developers and industry engineers had their eyes on the Soyuz rocket and their ears on the countdown. Precisely at the defined time, the rocket was launched, with the roaring sound of the burning fuel.
For a space mission, the launch is a point of convergence, a single point in space and time, where efforts of sometimes decades of development and work of many people of different backgrounds and locations are brought together. A space mission like Gaia combines so many challenges, that it can be considered a jewel of our modern, highly technological society. Gaia is one of the most ambitious missions of the Astrophysics programme of the European Space Agency to date.
After the commissioning period which lasted six months, the scientific operation of the Gaia mission has been handed over to the Data Processing and Analysis Consortium (DPAC), see~\cite{2008IAUS..248..224M}. The task of DPAC is to iteratively calibrate and reduce
the raw data, and to  deliver them to the scientific community
(see Sect.\ \ref{Section:ReleaseScenario} for a possible scenario of data releases).

\section{Gaia in a nutshell}

We summarise only very briefly\footnote{We already described the mission in \cite{Eyer2013}, issued before the launch of Gaia; another description can be found in \cite{deBruijne2012}. Updated information about Gaia can be found at the ESA website \url{http://www.cosmos.esa.int/web/gaia/}.} 
some key points of this mission: Gaia is a cornerstone mission of the European Space Agency. ESA selected EADS Astrium (now Airbus) as prime contractor to build the spacecraft. Gaia's goal is to observe all objects brighter than $G=20$\,mag,\footnote{The actual limit is currently set fainter, but may be subject to changes. See \cite{Jordi2010} for photometric transformations of the G, BP and RP bands.} including the bright objects down to magnitude $2$--$3$. The bright limit was initially set to $6$\,mag, however studies showed that brighter objects can be reached \citep{2014SPIE.9143E..0YM}. The Gaia spacecraft is continuously scanning the sky in TDI (Time Delay Integration) mode and performs astrometric, photometric, spectrophotometric and spectroscopic measurements. The nominal survey will last five years, with possibly one year of extension.

The processing of Gaia data is a challenge of the highest order and the DPAC consortium has started this tremendous task. DPAC is subdivided into nine Coordination Units (CU), namely: System Architecture (CU1); Simulations (CU2); Astrometric Core Processing (CU3); Object Processing (CU4); Photometric Processing (CU5); Spectroscopic Processing (CU6); Variability Analysis (CU7); Astrophysical Parameters (CU8); Catalogue Access (CU9).

\section{News on Gaia}
After the successful launch, Gaia was injected into its orbit around L2 (1.5 million km from Earth) with near-perfect precision. After that, a several-month commissioning phase started, which terminated at the end of July 2014.
Globally the spacecraft is functioning well, though several unexpected issues have
been identified. They are published on the Gaia webpage\footnote{See the webpage
\url{http://www.cosmos.esa.int/web/gaia/news} under 29/07/2014 ``The end of commissioning: Gaia starts routine operations''.}.
In summary:
\begin{itemize}
  \item Gaia is fainter than 
  expected, as seen from the ground: in order to attain Gaia's
  intended astrometric precision, one needs to know very precisely the position and velocity of
  the spacecraft (at precisions of about 150\,m and 2.5\,mm/s). 
  The Ground-Based Optical Tracking
  \citep[GBOT, see][]{2014SPIE.9149E..0PA} organises the observations of the spacecraft
  with optical telescopes. Before Gaia launch, GBOT performed optical ground-based 
  tests taking the Planck spacecraft as a proxy for Gaia. For example, the 
  1.2 m Euler telescope at La Silla and its twin 1.2 m Mercator telescope at La Palma performed
  such observations. The fact that Gaia is fainter than foreseen (about 21 mag instead of
  18 mag), implies that only larger 2m-class telescopes can be used for this task.
  
  \item There is varying stray-light on the focal plane. The reason for the
  stray-light is understood to have two main components, one is the light of the
  Milky Way and the other one originates from the Sun (fibres at the edge of the sun shield diffuse
  sunlight). Not much can be done to remove this. However, mitigation steps are being
  implemented\footnote{\label{footnote_label4Gaiaweb}For further details, see the ESA website 
  \url{http://www.cosmos.esa.int/web/gaia/news}.}
  to minimise the impact. Effectively, the stray-light degrades the performance at the faint end.
  The processing will be more complicated than anticipated since for a given
  object, one transit may not be much affected by stray-light, but another may. In other
  words, heteroscedastic data analysis procedures should be used.
  An important impact of the stray-light is on the Radial Velocity Spectrometer (RVS) 
  instrument:
  the expected performance at a given magnitude is
  shifted to $1$--$1.5$ magnitude brighter.
  The revised performances are described in Sect.\ \ref{Section:Performance}. One positive outcome is
  that {\it all} spectra will be taken at high resolution.

  \item The basic angle of $106.5^{\circ}$ (angle between the two field of views) has larger variations than
  predicted, according to the on-board Basic Angle Monitor. Studies indicate that it correctly measures
  the basic angle variations, and calibration work now focusses on achieving the highest possible accuracy.
  
  \item It seems that some water vapour escapes possibly from the service module and finds its way to the payload, causing contamination on optical parts, including the mirrors. Several decontamination campaigns were done, by heating the mirrors to get them clear again.  The level of the contamination is monitored and new decontamination campaigns might be performed if deemed necessary.
  
  \item The rate of micrometeoroid hits is larger than predicted from models. There is the need to modify the
  input model to match the observations. This will then reveal if there is a discrepancy in the micrometeoroid density,
  velocity distribution or impulse transfer function to Gaia at impact. These hits introduce discontinuities in the attitude of the satellite. They should be identified and taken into account in the reduction. Note that one reason for the improvement of the new Hipparcos reduction was due to such identifications \citep{2007ASSL..350.....V}.
\end{itemize}

We are grateful to ESA, the project scientist (Timo Prusti), the Gaia Science Team\footnote{GST is composed by: A.Brown, C.Jordi, S.Klioner, L.Lindegren, F.Mignard, T.Prusti, M.-S.Randich, C.Soubiran and N.Walton.}
(GST), and the Data Processing and Analysis Consortium Executive\footnote{DPACE is composed by: A.Brown, A. Vallenari, H.Siddiqui, C.Babusiaux, U.Bastian, D.Pourbaix, F.van Leeuwen, P.Sartoretti, L.Eyer, C.Bailer-Jones, X.Luri and V.Valette.} (DPACE) for their transparent 
attitude towards the problems that have appeared. 

While a list of problems may look somehow disheartening, the bottom line is that
the spacecraft is fully operational, and the mission is delivering and will deliver
fantastic scientific results on a very broad range of astrophysical topics.

\subsection{The currently estimated performance}
\label{Section:Performance}
After the spacecraft has become operational and started taking real data, the performance estimates have been revised. However, to have more reliable performance estimates, we should still wait until the calibration procedures are optimised.

The astrometric performance, given in terms of the mean parallax error at the end of the
mission, depends on the brightness of the considered object and its colour. The table in
Figure~\ref{fig:astroperf} is taken from the ESA Gaia webpage and shows a summary of the
astrometric performance measures. It should be noted that, on average, the errors on the position
and on the proper motion of the stars at the end of the mission can be derived from their
parallax error by multiplying the numbers given in Figure~\ref{fig:astroperf} by a factor
of 0.7 and 0.5, respectively.

\articlefigure[scale=.445]{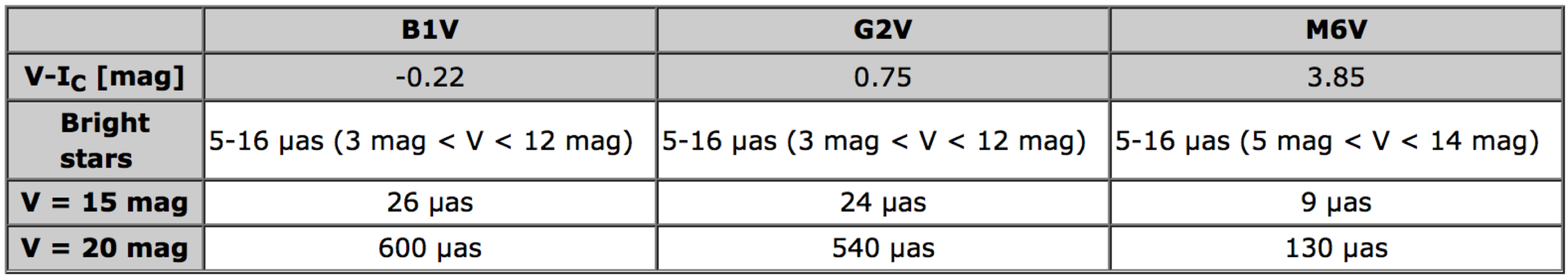}{fig:astroperf}{Astrometric performance summary as
found on the \href{http://www.cosmos.esa.int/web/gaia/science-performance}{Gaia ESA website} (December 2014).
These numbers describe the mean (sky average) parallax error at the end of the mission.} 

The performance of the Radial Velocity instrument is taken from the same webpage and is
reproduced in Figure~\ref{fig:RVSperf}.

\articlefigure[scale=.3]{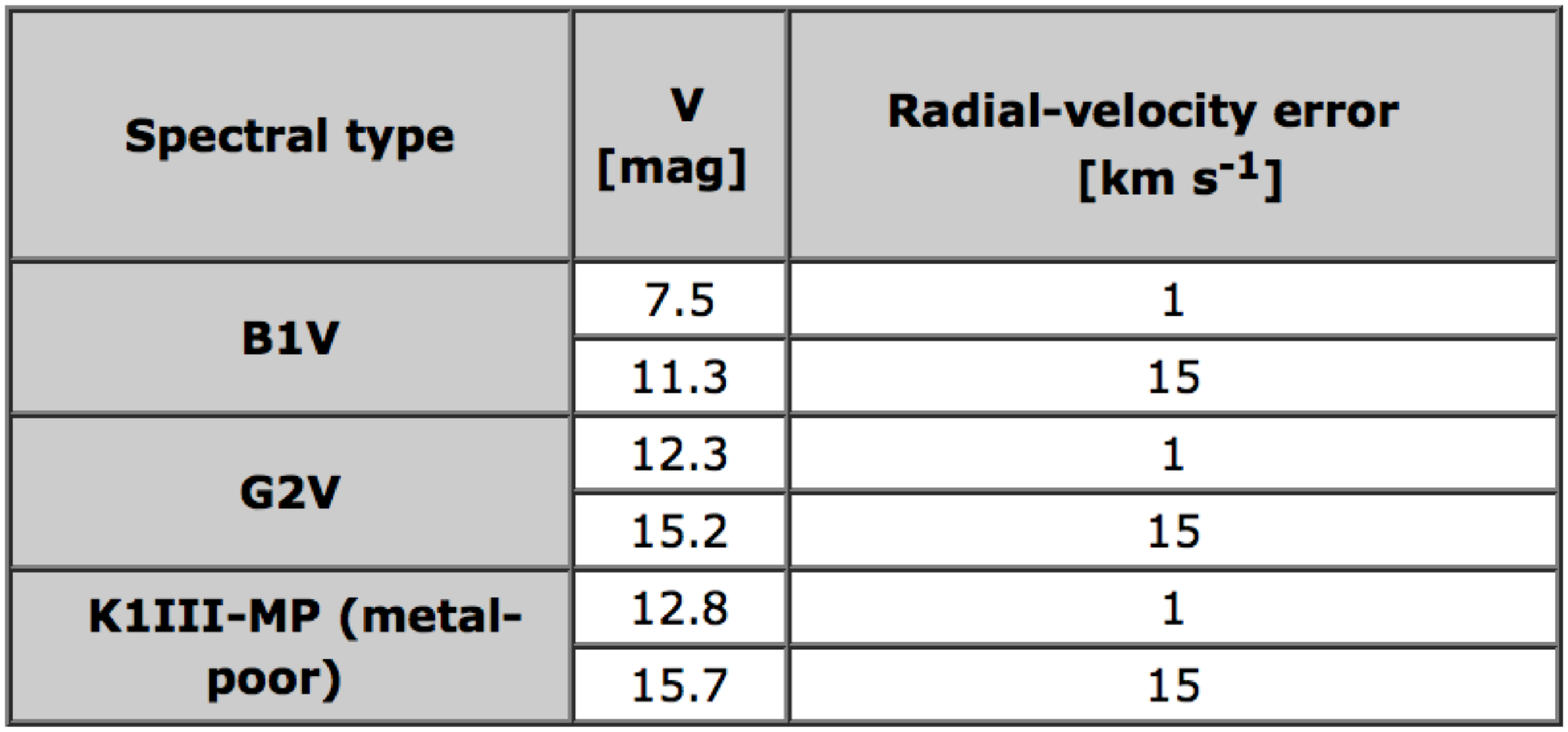}{fig:RVSperf}{Radial velocity performance summary
as found on the \href{http://www.cosmos.esa.int/web/gaia/science-performance}{Gaia ESA website} (December 2014).}

For the photometric performance, we will focus here on the $G$-band magnitude obtained
by the average of nine CCDs, which is called per-transit photometry. On average, there
will be $70$ such transits per source during the five-year mission (this estimate
includes $20\%$ of dead-time). The estimated uncertainty of the $G$-band measurements as
a function of magnitude is shown in Figure~\ref{fig:GmagErr}. We see that at bright
magnitudes there is a sawtooth behaviour due to the gating system which allows us to
observe bright stars (only a fraction of the CCD is integrated). We also remark a
discontinuity at magnitude $G=16$, which is caused by the change of the windowing
scheme (yet to be optimised).

\articlefigure[scale=.4,angle=270]{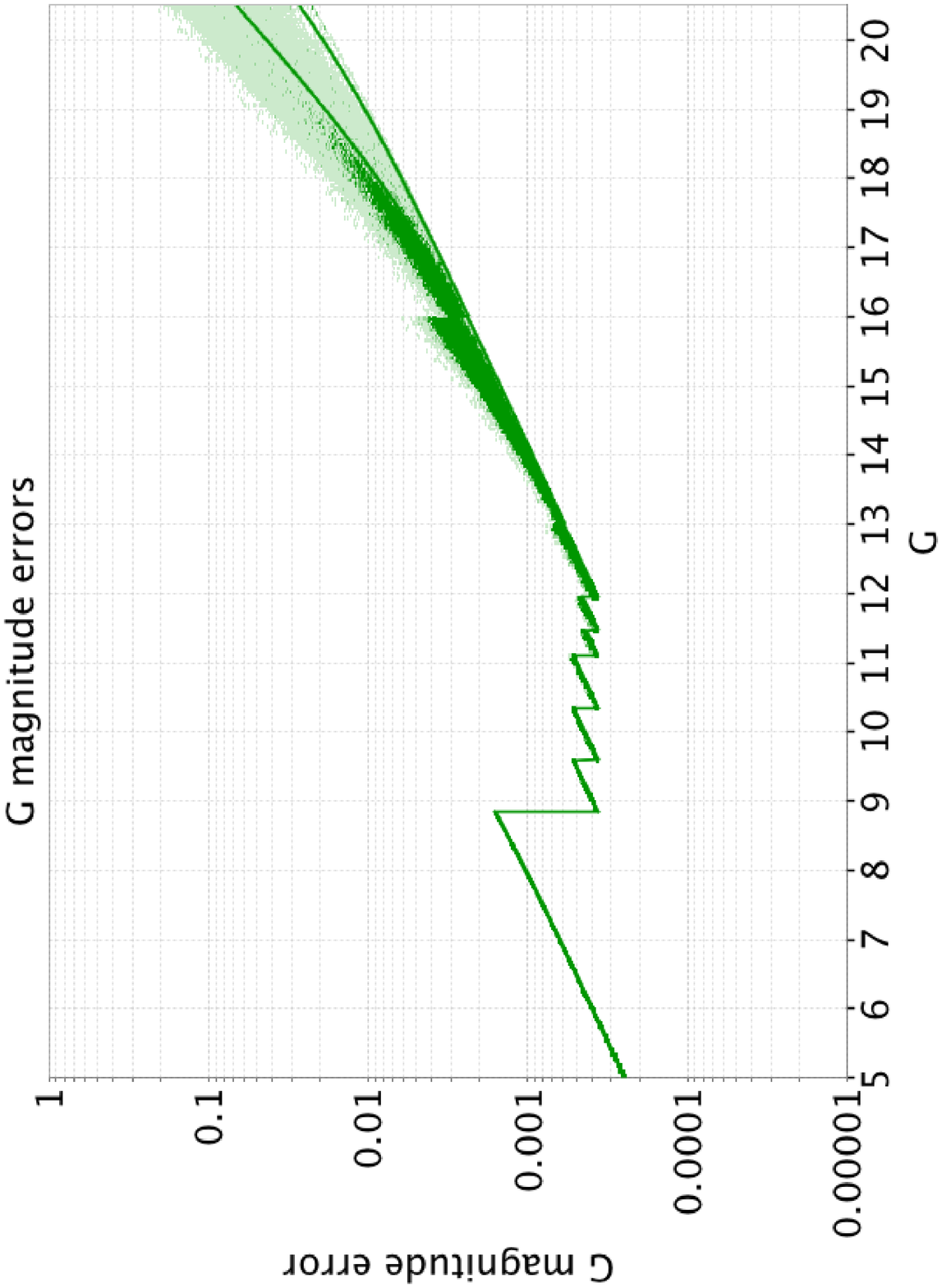}{fig:GmagErr}{$G$ magnitude uncertainty of the per-transit photometry as a function of $G$ magnitude, based on simulations, which account for stray-light.
At the faint end the bottom line indicates the pre-launch expected uncertainties, while the upper line indicates the mean in-orbit performance. The spread is caused by varying degrees of stray-light affecting the observations.}

\subsection{First Gaia results}

After the end of commissioning, the spacecraft has first started observing the sky in a
special observing mode, the Ecliptic Pole Scanning Law (EPSL), for a duration of $28$ days. After that, the spacecraft switched to observing the sky in the Nominal Scanning Law (NSL). 

Several of the DPAC Coordination Units are currently in operation (the initial data treatment, the first look, the astrometric processing for CU3, the photometric processing for CU5, the spectroscopic processing for CU6), while the others are using the calibrated data for their Operation Rehearsals (CU4, CU7 and CU8). Formally these other Coordination Units will start operation later.

We present here some results of these early data which were made public.

\subsubsection{The Ecliptic Pole Scanning Law}

In the EPSL mode, the axis of rotation of Gaia remains in the ecliptic plane, while the satellite spins at a rate of $4$ revolutions day$^{-1}$. The result is a regularly repeating observation of a limited number of sources. The aim of this limited-duration observing mode is to help start the photometric calibration. At the end of the $28$-day period, a transition to the Nominal Scanning Law (NSL) was performed, in which the satellite axis of rotation started precessing around the Sun at an angle of $45^{\circ}$, with a period of $63$ days, causing a more uniform coverage pattern over the whole sky. This NSL mode is optimised to provide observations that best constrain the astrometric parameters that will be derived from them.

Figure~\ref{fig:epsl-28d} shows the maximum number of observations expected for sources which
are always within the EPSL footprint (from the first to the $28$th day). The number of samples is maximal
($28$ days $\times$ $8$ transits per day $= 224$ transits) up to $1.44^{\circ}$ from the ecliptic poles
and then decreases to the minimum value ($8$ transits per day $\times$ the field-of-view width of
$0.69^{\circ} /$ the precession rate of the spin axis by almost $1^{\circ}$ per day $= 5$ transits) at the ecliptic
equator. A subset of sources within $3^{\circ}$ is sampled by over $100$ quasi-regular measurements,
making it not only very useful to start the photometric calibration, but also a remarkable set of multi-epoch observations.

Consecutive per-transit measurements are separated by:
\begin{itemize} 
  \item 0.074 day (about $1$h $46$m) from the succession of transits measured first by the preceding field of view and then by the following field of view.
  \item 0.176 day (about $4$h $14$m) from transits measured first by the following field of view and then by the preceding one.
\end{itemize}

When considering the per-CCD photometry,  the number of observations has to multiplied by about a factor~$9$. The time separation between successive CCD-measurements is then shorter and reduces to about $4.9$~seconds.

\articlefigure[scale=.52]{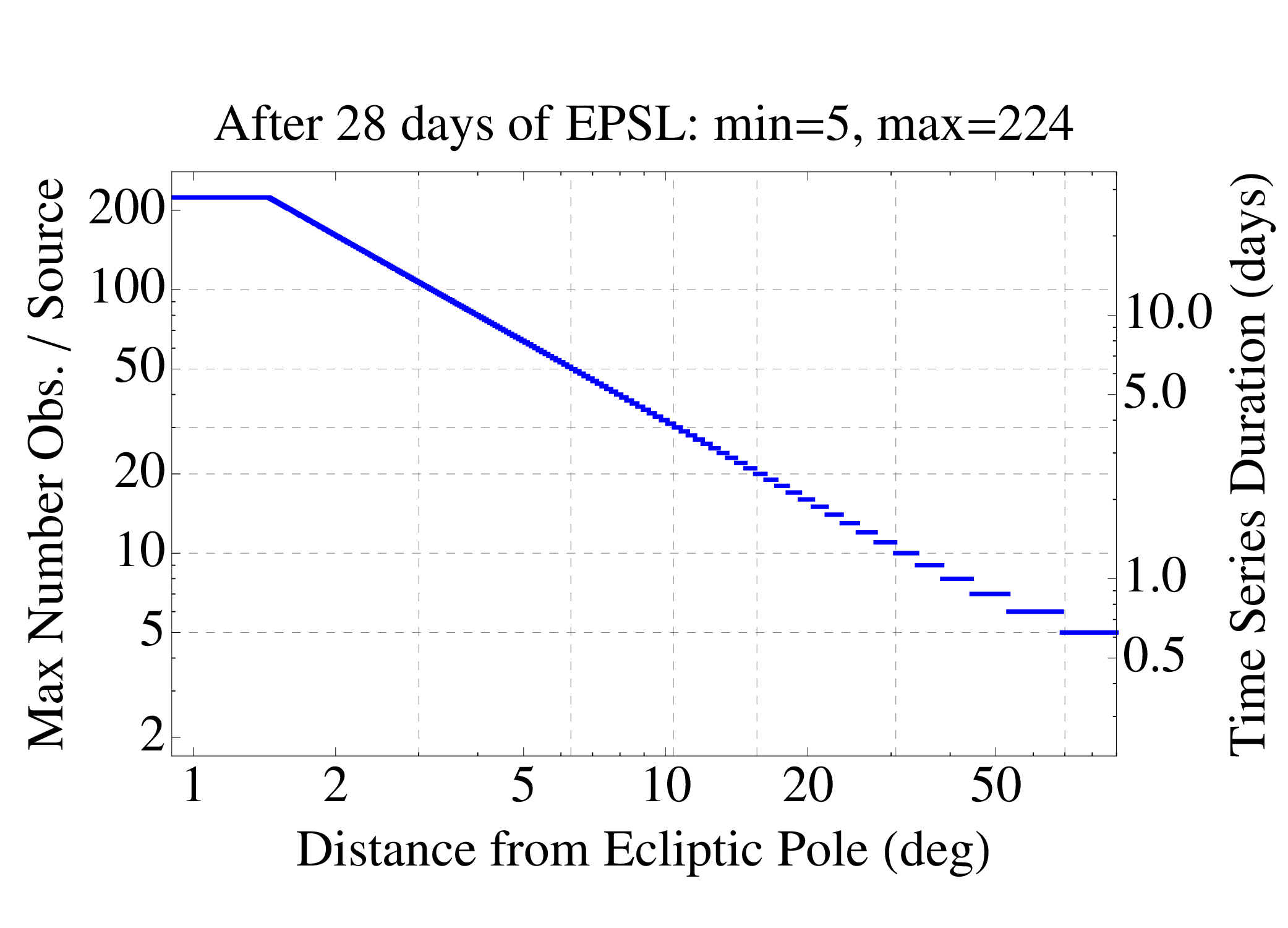}{fig:epsl-28d}{The maximum number of observations for sources covered by the EPSL footprint is presented as a function of angular distance from the ecliptic poles.}

\subsubsection{Results from the 28 days of EPSL}
As there is a strict data policy by ESA on Gaia data, without proprietary data for DPAC, all the results presented here can be found on the ESA Gaia website.

Within the consortium, the temptation was great to check some first results with respect to known variable stars. Carme Jordi from Barcelona University looked for matches in the Hipparcos catalogue and found an eclipsing binary, GW\,Dor, which was observed by Gaia. Although the data was not calibrated, the results were very encouraging and are presented in Figure~\ref{fig:GWDor}.
\articlefigure[scale=.40,angle=270]{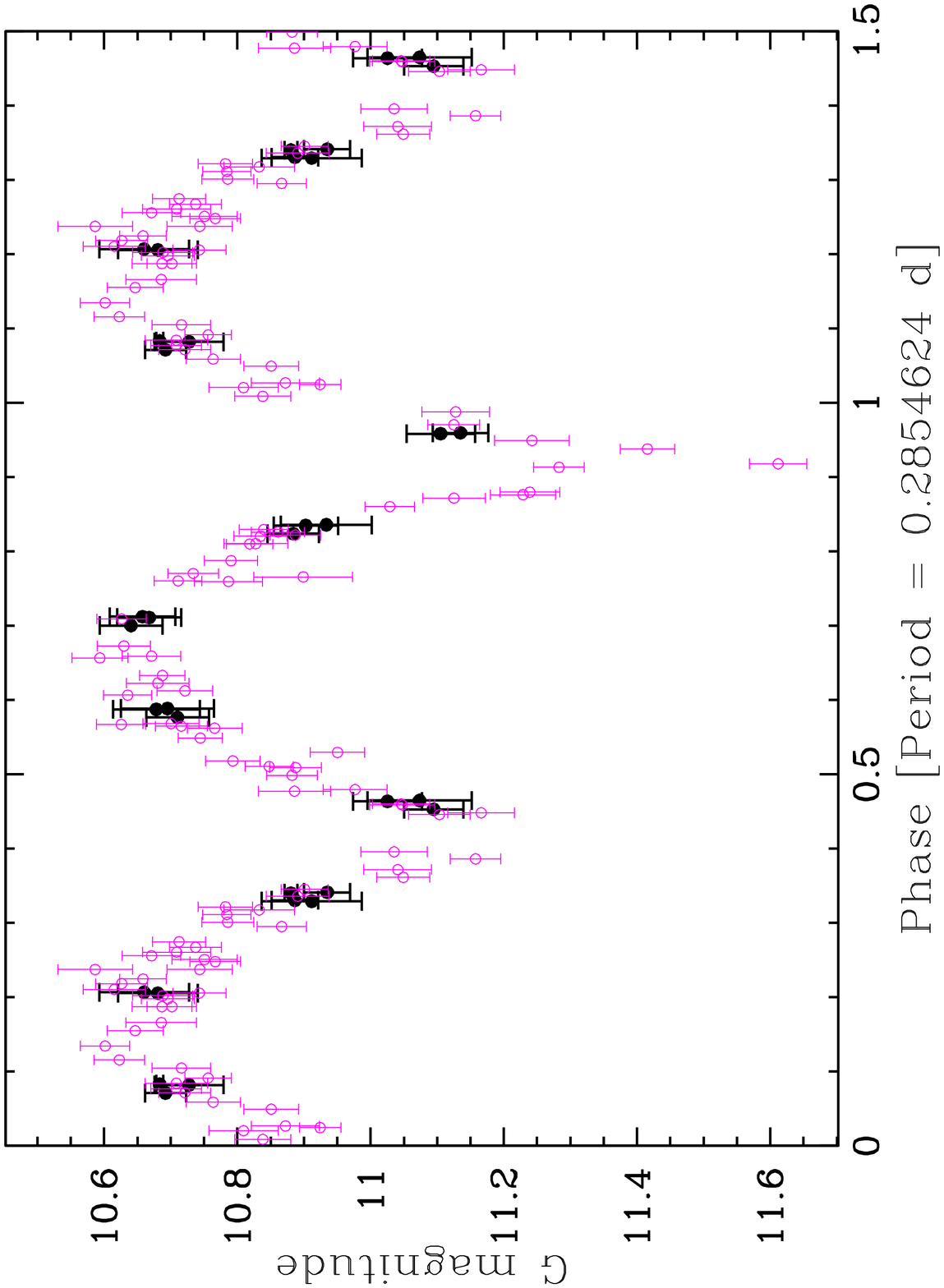}{fig:GWDor}{The first presented variable star of Gaia: GW\,Dor, an eclipsing binary. Gaia data (black filled symbols) are compared with Hipparcos data (magenta/grey open symbols) (data courtesy of C.Jordi).}

During a CU7 meeting in November 2014, D.~W.~Evans  presented several variables detected by Gaia, including a surprising microlensing candidate (see Figure~\ref{fig:microlensing}). The duration of the event seems short and the amplification seems low. We will check if this source does not have other ``bursts'', and if 
other events of this kind are present in the data. This case is shown with per-CCD photometry: the number of measurements is large, more than $1000$ measurements over $28$ days. From this case, we learn that the photometry of Gaia is performing as expected. There are some outlying values and probably some other unpleasant features in the data, but over the coming years the data reduction will be corrected and improved.

\articlefigure[scale=.40]{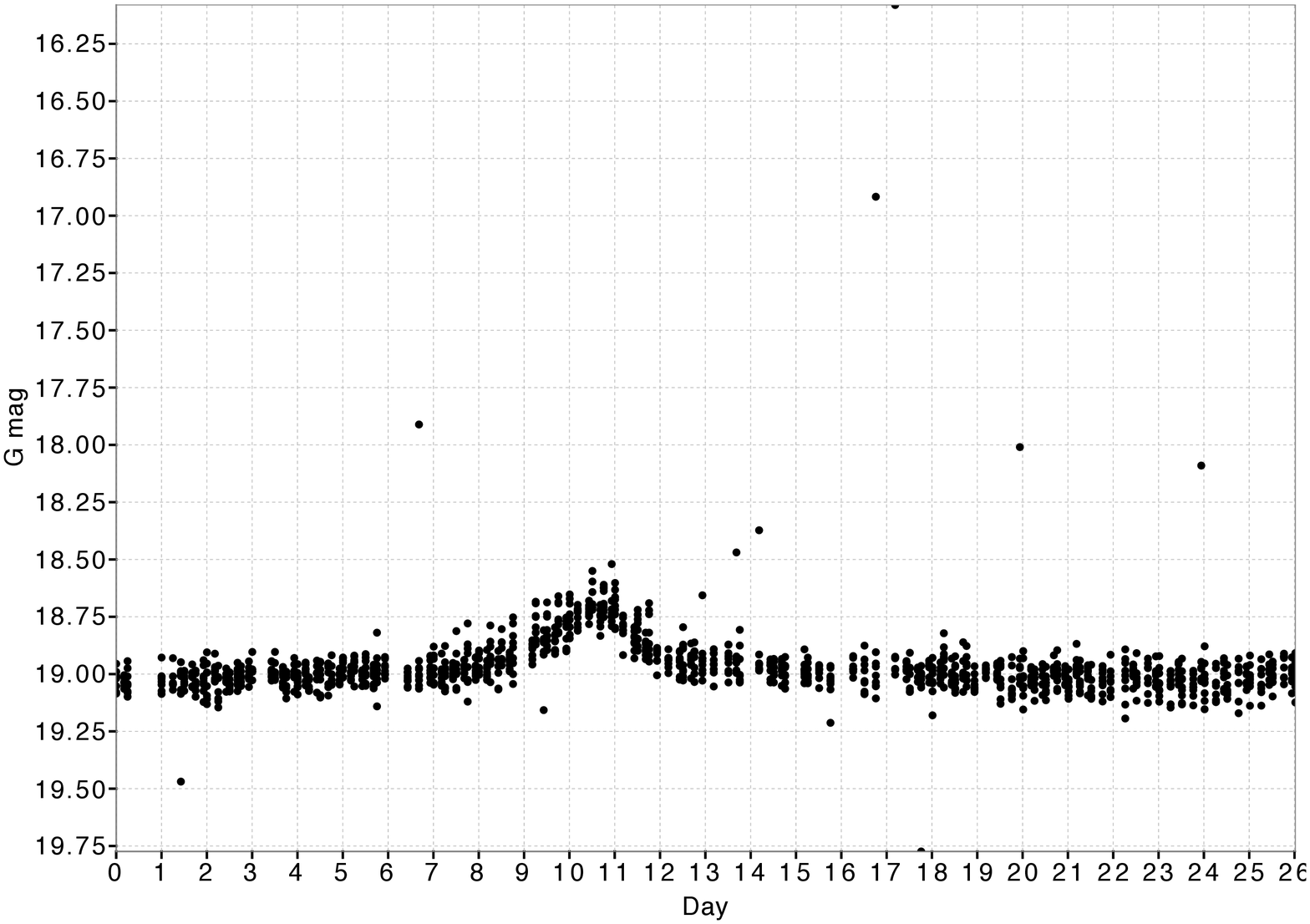}{fig:microlensing}{A microlensing candidate (per-CCD photometry).}

\subsubsection{The Science Alert System}
A Gaia photometric alert system is operated by DPAC at the Institute of Astronomy in Cambridge \citep[see][]{2013RSPTA.37120239H}.
The first alert was published in August 2014 and was a type Ia supernova (see Figure~\ref{fig:supernova}).
\articlefigure[scale=.80]{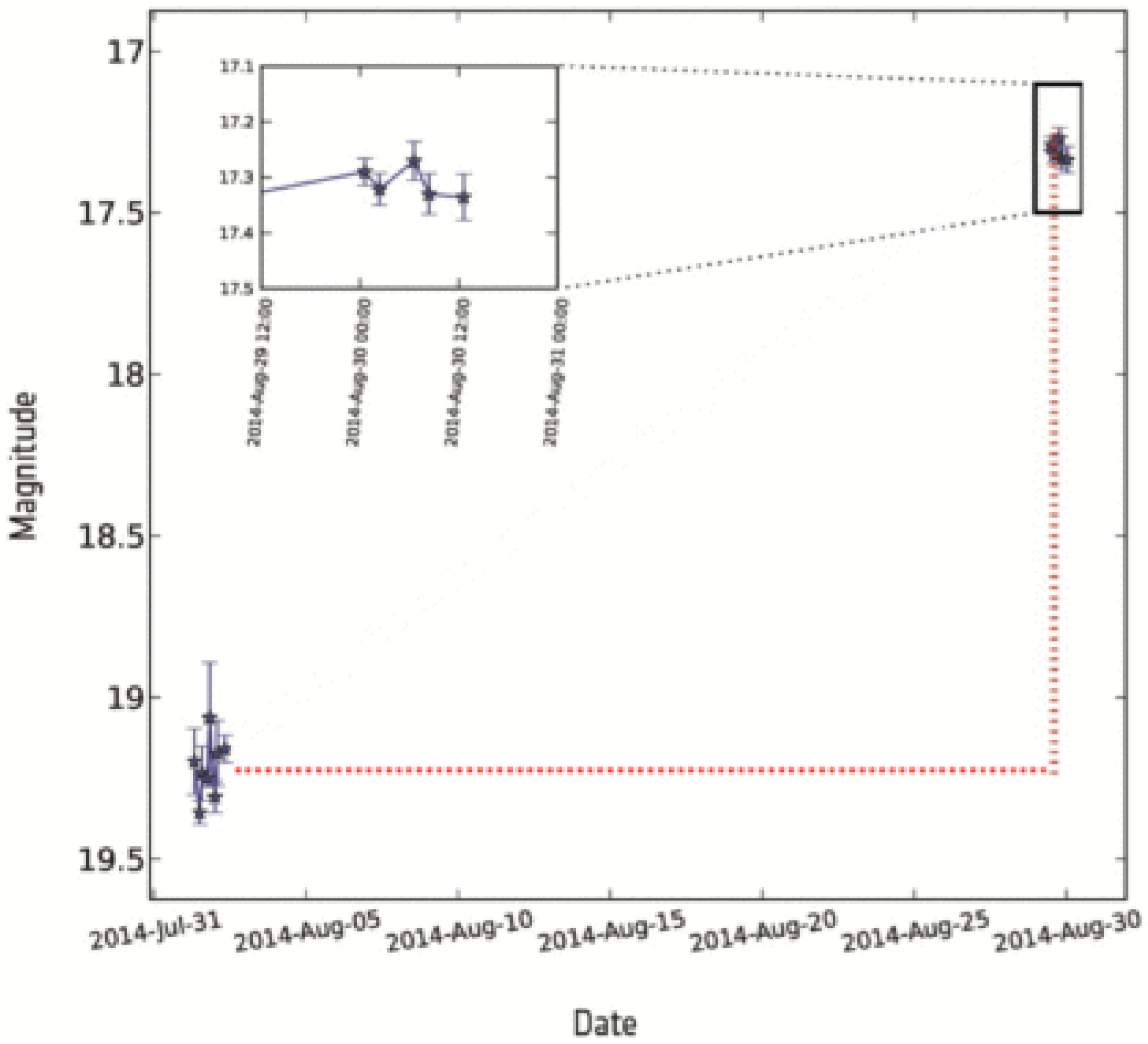}{fig:supernova}{The first type Ia supernova detected by Gaia, from
\href{http://www.esa.int/Our_Activities/Space_Science/Gaia/Gaia_discovers_its_first_supernova}{ESA press release (12/09/2014)}. Figure modified by L.Eyer.}
The alerts are currently announced on a Cambridge
University webpage\footnote{\url{http://gaia.ac.uk/selected-gaia-science-alerts/}}. The Alert system is functioning in verification mode and has produced $97$ alerts (as of December 16, 2014), among which seven are supernovae. Excitingly, an AM~CVn star was also discovered in outburst. Amateur astronomers first noticed that the object is also eclipsing (Campbell et al., in prep). Such systems are very rare, and this one is only the third known object of this kind!  Gaia offers a unique combination of high spatial-resolution photometry with a stable PSF, at a well defined
observing cadence. Importantly, Gaia also measures near-simultaneous low-dispersion spectrophotometry. This is a significant advantage over ground-based transient surveys, where follow-up classification spectroscopy is usually overwhelmed by the transient discovery rate. 
The selection of objects to follow-up, i.e., to expend valuable resources on, is necessarily biased (it depends on human interaction) and is significantly incomplete. 
Therefore, a spectroscopic classification method has been developed \citep{2014MNRAS.442..327B}, using the Gaia spectrophotometry. In the longer term, the team plans to explore the spectrophotometry to look for unusual signatures independently
from variations in flux. Throughout the mission, Gaia's typical observing cadence (measurements separated by hours and then months) does mean that follow-up data are still needed to fully exploit the discoveries. In conclusion, the expected harvest of Gaia looks really very promising.

There is also an alert system for solar system objects, entitled Gaia-FUN-SSO. Alerts are triggered on sources detected by Gaia, not identified as  
star, galaxy, or known solar system object (SSO). Known SSOs are processed every six months in the long-term chain (SSO-LT). These alerts are, however, processed daily in the so-called short-term chain (SSO-ST). Indeed, new sources will cross Gaia fields once or twice, seldom more
often,
providing a limited orbital arc of a few hours only. Such data are not sufficient to uniquely determine the heliocentric orbit of the newly detected objects. As a result, 
apparently 
``new'' detections of the same object could occur on multiple occasions. This would put a higher burden on the SSO-LT, which attempt to link together observations of unidentified sources. Follow-up observations of these detections from Earth are thus required to characterize their orbits. The SSO-ST chain will therefore provide daily the list of new sources detected by Gaia, together with indications to observe them from Earth (i.e., the object ephemeris). At the time of this writing, the SSO-ST and alert system are not routinely operating yet. The first alerts are expected to be triggered starting from April 2015. They will be publicly disseminated on the alert portal \url{http://gaiafunsso.imcce.fr/}
(registration required). Further information can be found on the internet (see the wiki at \url{https://www. imcce.fr/gaia-fun-sso/} and the Gaia-FUN-SSO workshop web site at \url{http://www.imcce.fr/hosted_sites/gaiafun2014/}). A telescope network (Gaia-FUN-SSO) has been set up and is organized by the IMCCE, Paris Observatory (Thuillot et al. 2014). The network ensures a world-wide coverage of telescopes ready to dedicate a fraction of their observing time to observe the alert in a rapid-response fashion. After detection by Gaia, there will be a delay of about 48h before the alerts are triggered during which the data are stored on-board, downlinked to Earth, pre-processed, sent to the DPCC in Toulouse, processed, and finally sent to the IMCCE for ephemeris generation and dissemination. Owing to the very limited arc of orbit observed by Gaia, the uncertainty on the position of new objects will increase with time. Observers will therefore have only one 
to a few nights to attempt finding the sources. Over the past couple of years, this network performed several test campaigns to observe about 10 solar system objects, among which the near-Earth asteroid (99942) Apophis (Thuillot et al. submitted). 

\section{Binaries and exo-planets}
The realm of binary stars covers a very broad range of topics in astrophysics:
indeed, multiplicity is a basic phenomenon of stellar formation, and binaries can be
used to test the theory of stellar evolution, especially when their components are also
probed through asteroseismology \citep{Huber2014}. They can also serve as standard
candles, provided the effective temperatures of their components are determined.
Furthermore, they play an important role in the dynamics of stellar systems, such as
globular clusters. The ever growing detection rate of exo-planets is boosting the study
of two-body and multiple systems.

\subsection{Binaries: back to basics}
The wealth of information which can be derived from binary stars is the most important
source for measurements and estimates of astrophysical parameters of stars, mainly the
radius and mass. In Tables~\ref{Tab:spectoscopicB}, \ref{Tab:astrometricB} and
\ref{Tab:eclipsingB}, we provide a reminder of what can be derived from observations
on the orbital and astrophysical parameters of binary stars. We consider here binary
systems with components of radius $R_1$ and $R_2$, of mass $m_1$ and $m_2$, with an
orbital period $P$, eccentricity $e$, inclination $i$, and semi-major axes $a_1$ and
$a_2$.

In Table~\ref{Tab:spectoscopicB}, we have considered the two usual categories of
spectroscopic binaries: SB1 for those where only the spectrum of one component is
visible, and SB2 for those where the spectra of both components can be identified.
SB1 and SB2 are purely spectroscopic, i.e. we assume that they are too tight for their
orbit to be measurable through astrometry. Regarding the stellar parameters, in the SB1
case, only the so called {\it mass function} can be derived from the radial velocity
data: 
$$
f(m_2) = \frac{(m_2 \sin i)^3}{(m_1+m_2)^2}
$$
In the SB2 case the mass ratio and a lower limit to each mass can be derived. 

\begin{table}[!ht]
\caption{\label{Tab:spectoscopicB} Spectroscopic Binaries.}
\begin{center}
{\small
\begin{tabular}{llcl}  
    Type       & Observables & Physical laws & Derived parameters\\ \hline
SB1 &$P_\mathrm{orb}$ \& $v_1\sin i=f(t)$&Kepler 1,2,3& $e$, $a_1\sin i$, $\frac{(m_2\sin i)^3}{(m_1+m_2)^2}$\\
    &                                   &                     &               \\ \hline
SB2 &$P_\mathrm{orb}$, $v_1(t) \sin i$  &Kepler 1,2,3         &$e$, $a_1\sin i$, $a_2\sin i$, \\
    &\& $v_2(t) \sin i$                 &+ centre of mass def.&$m_1 \sin^3i$, $m_2 \sin^3i$\\
    &                                   &                     &$q\equiv\frac{m_2}{m_1}=\frac{v_1}{v_2}$\\ \hline
\end{tabular}
}
\end{center}
\end{table}

In Table~\ref{Tab:astrometricB}, we list the various cases of the astrometric binary
category. Astrometric binaries are objects for which Gaia can trace the orbits
projected on the plane of the sky. We have considered in the Table four sub-categories: those which benefit from Radial Velocity (RV) measurements (because they are bright
enough to be observed spectroscopically) and those which do not, each of these two
classes being split into binaries whose parallax is known, and those without a
significant parallax. Although it might appear surprising at first sight, we should
expect that some astrometric binaries will have insignificant parallax. In such
systems, only the primary is visible because of a low mass ratio, but it is bright
and massive enough that it can be seen at large distance; the orbit remains wide enough
to subtend a measurable angle, but tight enough that the orbital period remains of
the order of the Gaia lifetime.

If the absolute orbits of both components are accessible as well as
the parallax, the full solution of the observed orbits provides the masses of the
individual components. 

\begin{table}[!ht]
\caption{\label{Tab:astrometricB} Astrometric Binaries. Orbital and stellar parameters
that can be derived with astrometry when absolute orbits are measured. The cases
without a significant parallax are envisaged. Note that the 3rd line (2 visible comp.,
no parallax) is a case that will not occur in the Gaia context; it is included only
for the sake of completeness.}
\begin{center}
{\small
\begin{tabular}{llcl}  
    Type              & Observables                       &Physical laws& Derived parameters          \\ \hline
1 visible comp.,      &$P_\mathrm{orb}$, size and shape   & Geometry    &$i$, $a_1["]$, $e$,          \\
no parallax           &of projected orbit,                &\& Kepler 1,2&$\omega$, $\Omega$, $T_0$    \\ 
                      &\& $v_1(t)\cos i$                  &             &                             \\ \hline
{\sl idem} + parallax & {\sl idem} + $\varpi_1$           &+ Kepler 3   &+ $a_1[A.U.]$,               \\
                      &                                   &             &$\frac{m^3_2}{(m_1+m_2)^2}=\frac{a^3_1}{P^2_\mathrm{orb}}$\\ \hline
2 visible comp.,      &$P_\mathrm{orb}$, size and shape   & Geometry    &$i$, $a_1["]$, $a_2["]$,     \\
no parallax           &of projected orbit,                &\& Kepler 1,2&$e$, $\omega$,  $\Omega$,    \\ 
                      &$v_1(t)\cos i$, $v_2(t)\cos i$     &             &$T_0$, $q=\frac{a_1}{a_2}$   \\ \hline
{\sl idem} + parallax & {\sl idem} + $\varpi$             &+ Kepler 3   &+ $a_1[A.U.]$, $a_2[A.U.]$,  \\
                      &                                   &             &$m_1$, $m_2$                 \\ \hline
With RVs but          &+$v_1(t) \sin i$, $v_2(t) \sin i$  &Kepler 1,2,3 & {\sl idem as} above,        \\
no parallax           &                                   &             & but more                    \\
                      &                                   &             & precise \& accurate         \\ \hline
With RVs              & {\sl idem} + $\varpi$             &Kepler 1,2,3 &{\sl idem}, but even more    \\
 + parallax           &                                   &             & precise \& accurate         \\ \hline
\end{tabular}
}
\end{center}
\end{table}

Table~\ref{Tab:eclipsingB} presents the case of eclipsing binaries, which have the
advantage of yielding the radii in addition to the masses, if RVs are available.
In general, systems with only partial eclipses will not allow us to determine the
individual radii -- at least not in a precise and unambiguous way -- but they do provide
a robust estimate of the sum of radii, whether absolute (if both RVs are available) or
relative to the orbital semi-major axis. If only one RV is measured, one gets the mass
function, but without the $\sin^3i$ uncertainty because $i$ is measurable through the
eclipses. If an eclipse is total (the other one being a transit) and both RVs are
measured, then accurate radii and masses can be determined, giving the opportunity to
estimate to the stellar densities. Paradoxically, the inclination $i$ may be less
accurate and precise in case of total eclipse than of partial one, but this is of
little importance since $i$ is very close to $90^\circ$ for totally eclipsing systems.

\begin{table}[!ht]
\caption{\label{Tab:eclipsingB} Eclipsing binaries, assuming they are too tight for astrometric orbit measurements. Information from photometry alone is
outlined, as well as the additional information given by radial velocity data.}
\begin{center}
{\small
\begin{tabular}{llcl}  
    Type               & Observables                       &Physical laws        & Derived                          \\
                       &                                   &                     & parameters                       \\ \hline
Partial eclipses       &$P_\mathrm{orb}$, light curve shape&Geometry \& Kepler 1,2&$i$,                             \\
\& photometry          &                                   &                     &$e\cos\omega$,                    \\
                       &                                   &                     &$\approx e\sin(\omega)$,          \\
                       &                                   &                     &$(R_1+R_2)/a$                     \\ \hline
         + RV, SB1     &+ $v_1(t) \sin i$                  &+ Kepler 3           &$a_1$, $\frac{m^3_2}{(m_1+m_2)^2}$\\ \hline
         + RV, SB2     &+ $v_1(t) \sin i$, $v_2(t) \sin i$ &                     &$a_1$, $a_2$, $a$                 \\
                       &                                   &                     &$m_1$, $m_2$,                     \\
                       &                                   &                     &$(R_1+R_2)$                       \\
                       &                                   &                     &                                  \\ \hline
Total eclipse          &$P_\mathrm{orb}$, light curve shape,&Geometry \& Kepler 1,2&$i$,                              \\
\& photometry          &abs. mag. of eclipsing             &                     &$e\cos\omega$,                    \\
                       &component                          &                     &$e\sin(\omega)$,                  \\
                       &                                   &                     &$R_1/a$, $R_2/a$                  \\
                       &                                   &                     &$M_\mathrm{V1}$,  $M_\mathrm{V2}$ \\ \hline
         + RV, SB1     &$v_1(t) \sin i$                    &+ Kepler 3           &$a_1$, $\frac{m^3_2}{(m_1+m_2)^2}$\\ \hline
         + RV, SB2     &$v_1(t) \sin i$, $v_2(t) \sin i$   &                     &$a_1$, $a_2$,                     \\
                       &                                   &                     &$m_1$, $m_2$,                     \\
                       &                                   &                     &$R_1$, $R_2$,                     \\
                       &                                   &                     &$<\rho_1>$, $<\rho_2>$            \\ \hline
\end{tabular}
}
\end{center}
\end{table}

\subsection{Gaia and Binaries}

Binaries, multiple systems and more generally optical multiple stars have a significant impact on the measurements and data processing of Gaia.
Globally, the presence of such systems is bound to complicate the data processing. Within the consortium, these systems impact all of the
Coordination Units. Conversely, all instruments of Gaia will have a scientific impact on our knowledge of binaries. Indeed, the Gaia mission
covers all the categories of binaries mentioned above. DPAC has therefore been actively preparing, in a systematic way, the processing and analysis of the non-single stars. The general processing foreseen in CU4 is described in \cite{Pourbaix2011}, the processing of spectroscopic binaries with the RVS instruments is covered by
\cite{Damerdji2012}, and the case of eclipsing binaries by \cite{Holl2013} and by \cite{Siopis2012}.

The estimates made in \cite{Eyer2013} have not changed: namely, ``we find that $30$ million objects will be processed as astrometric non-single stars, $8$~million as spectroscopic binaries (of which $59\%$ will be SB2) and $4$~million as eclipsing binaries (of which $12\%$ will be spectroscopic binaries).''

Very soon the first results of Gaia will allow us to revise these estimates. We can guess
that the short period eclipsing binaries will be detected first. Thanks to the BP and RP
photometry and Radial Velocity instrument, these objects will be easily distinguished from
pulsating stars, which are often a source of confusion when single photometric band observations are used.

There are developments made to assess if it is possible to derive the Full Width 
at Half Maximum (FWHM) and contrast of the correlation function from the RVS instrument, which could be an informative diagnostic to further improve the identification of binary systems.

\subsection{Gaia and exo-planets}

One of the important science cases of Gaia is the astrometric detection of exo-planets.
Obviously, Gaia will provide parallaxes of the currently known exo-planet systems, thus
deepening our capabilities to study them. However, more importantly, it will serve
as an all-sky magnitude-limited exo-planet census, expanding the range of stellar
parameters of previous planet searches.  The current estimates indicate that Gaia will
detect astrometrically of the order of $20,000$ Jovian exo-planets
\citep{Perryman2014, Sozzetti2011}. To this date there has been no confirmed
astrometric detection yet.

Gaia is also expected to contribute to the growing population of transiting
exo-planets, although the time sampling of Gaia is sparse and it is far from being
optimal for this purpose. Thus, only fortuitous timing of the transits may result in
detection. Nevertheless, carefully timed complementary ground-based observations
should allow such detections \citep{Dzigan2011}. Using this technique,
\cite{Dzigan2012} estimate that Gaia will detect of the order of hundreds or a few
thousand exo-planets.

On top of detecting more planets through astrometry and photometry, Gaia will
contribute significantly to the understanding of planet formation and evolution
by precisely measuring the astrophysical characteristics of the planet-hosting
stars. Characteristics like effective temperature, stellar radius, age, kinematic
population membership etc. will serve to finely tune planet formation models and
greatly improve our understanding of this fundamentally important topic.

 \section{Release scenarios}
\label{Section:ReleaseScenario}
Currently (December 2014), five releases of the Gaia data are foreseen. We present a schematic view of these releases in Figure~\ref{fig:ReleaseScenario}. The precise content and the
exact dates of the releases are still under discussion and may be subject to change. The actual content of the releases and dates are obviously constrained by the data analysis, and
will be reviewed at several levels, by DPACE, GST, and ESA\footnote{In December 2014, an article \citep{2014arXiv1412.8770M} proposed to combine data from the Hipparcos satellite, the
Tycho data \citep{2000A&A...355L..27H}, with early Gaia data to provide early 5-parameters astrometric solution, thus including the parallax. We will see if such a proposal can converge into an early release.}. This (evolving) data release scenario can be found on the ESA website (\url{http://www.cosmos.esa.int/web/gaia/release}). This website will give the latest information on the different releases.

\articlefigure[scale=0.65,angle=180]{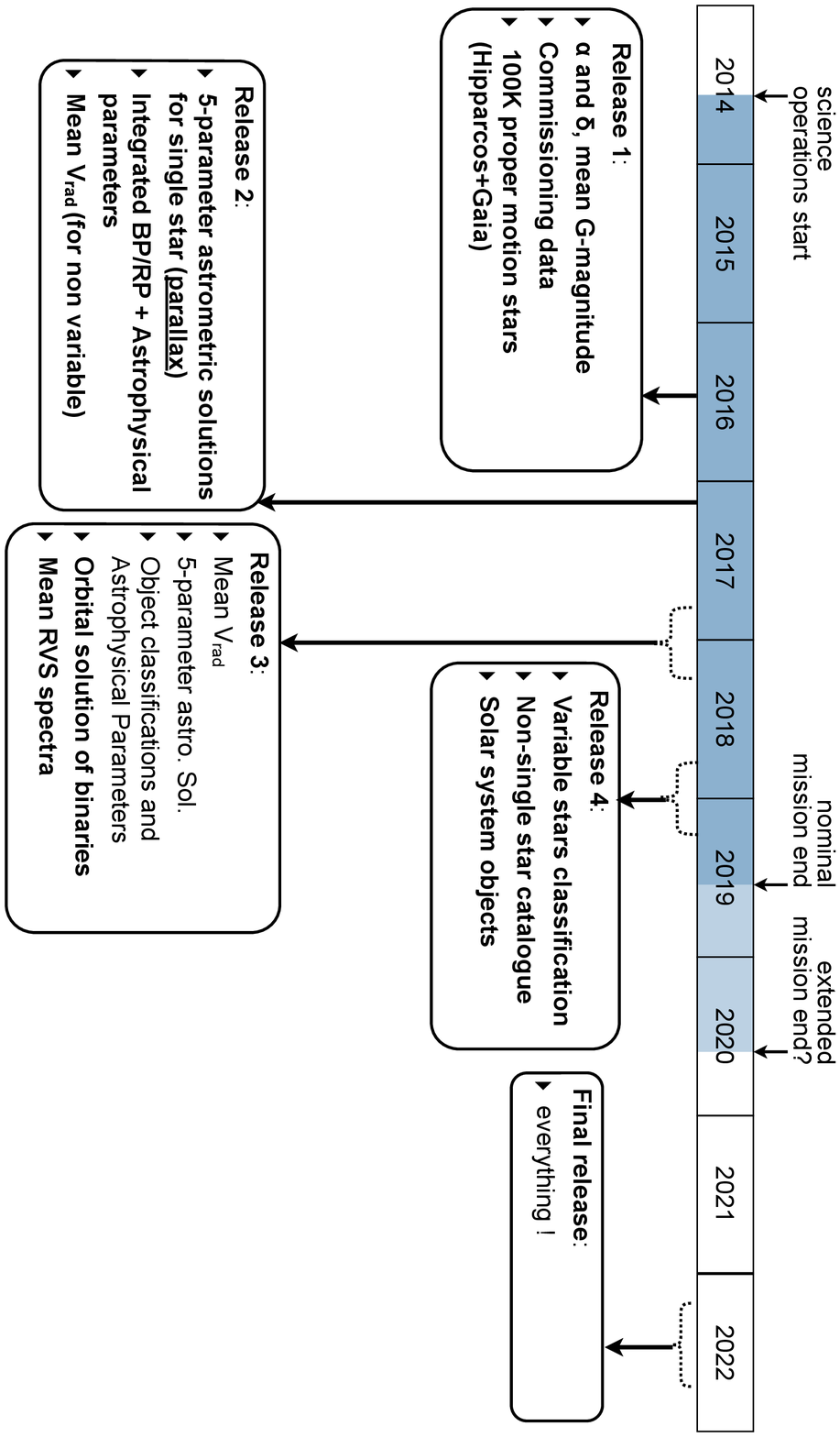}{fig:ReleaseScenario}{The release scenario: the five currently foreseen releases of the Gaia catalogue.}

We can anticipate the following:
\begin{itemize}
\item In the third data release, the orbital solutions of binary stars may be included, for binaries with periods from two months to $75\%$ of the observing time considered in the analysis.

\item In the fourth release, the catalogue of non-single stars could be delivered.

\item In the current plan, we remark that the results of the variability processing and analysis are published only in the fourth release. The variability information in the fourth
release aims at delivering the result of the global analysis of variable objects. We hope, however, that some of the variability results can be released earlier, for example, those
involving stars of specific variability types like RR\,Lyrae stars or short-timescale eclipsing binaries, as long as the estimates of completeness and contamination are well constrained
and considered at acceptable levels. For objects like RR\,Lyrae stars and short-period eclipsing binaries, some regions of the sky are quite well known, which therefore should allow us to readily estimate reliable contamination and completeness rates.
\end{itemize}

\section{Conclusions}
Gaia is a unique mission because it surveys the entire sky with a very diverse set of
instruments performing near simultaneous measurements of sources. It will provide
exquisite astrometry (positions, parallaxes and proper motions) together with
photometric/spectroscopic measurements, it includes the ``bright sky'', and it uses a time
sampling that is very particular and different from ground-based surveys.

Among many other different scientific topics, Gaia will study binary stars in unprecedented detail, by providing fundamental physical properties of the binaries and
their variability. The large numbers of binaries observed by Gaia will allow us to have
a dual approach: on the one hand, to describe the properties of binaries in
general (e.g. the distribution of the orbital parameters) and, on the other hand, to search for very rare objects. 

Gaia is not alone and certainly its data will prove even more interesting when combined with other data sets, complementing other projects such as LSST, OGLE, CoRoT,
Kepler, TESS, PLATO, CHEOPS, etc.
Finally, the follow-up of Gaia discoveries with ground-based facilities, already during
the mission lifetime, will be most fruitful employing large telescopes, as well as small
ones for the bright sources.

\acknowledgements We want to thank Timo~Prusti and Jos~de~Bruijne for their comments on the manuscript.
S.Z. acknowledges support by the Israel Ministry of Science and Technology, via grant 3-9082.


\end{document}